\def\aa{{A\&A}}

\def\annrev{{ARA\&A}}
\def\apj{{ApJ}}
\def\apjs{{ApJS}}

\def\mnras{{MNRAS}}

\documentclass[cup5b]{caps}
\usepackage{graphicx}
\usepackage{amssymb}
\usepackage{ociwsymp3}   

\HeadText{R. F. Mushotzky}

\begin{document}

\pagenumbering{arabic}

\author[]{RICHARD F. MUSHOTZKY\\ NASA/Goddard Space Flight Center}

\chapter{Clusters of Galaxies: An X-ray Perspective}

\begin{abstract}
There has been extensive recent progress in X-ray observations of
clusters of galaxies with the analysis of the entire {\it ASCA}
database and recent new results from {\it Beppo-SAX}, {\it Chandra},
and {\it XMM-Newton}.  The temperature profiles of most clusters are
isothermal from 0.05--0.6 $R_{\rm viral}$, contrary to theoretical
expectations and early results from {\it ASCA}. Similarly, the
abundance profiles of Fe are roughly constant outside the central
regions. The luminosity-temperature relation for a very large sample
of clusters show that $L_{\rm X}\propto T^3$ over the whole observable
luminosity range at low redshift, but the variance increases at low
luminosity, explaining the previously claimed steepening at low
luminosity. Recent accurate cluster photometry in red and infrared
passbands have resulted in much better correlations of optical and
X-ray properties, but there is still larger scatter than one might
expect between total light and X-ray temperature and luminosity. The
velocity dispersion and the X-ray temperature are strongly correlated,
but the slope of the relation is somewhat steeper than expected. The
surface brightness profiles of clusters are very well fit by the
isothermal $\beta$ model out to large radii and show scaling
relations, outside the central regions, consistent with a
$\Lambda$-dominated Universe.

At high masses the gas mass fraction of clusters is quite uniform and
is consistent with the low WMAP value of $\Omega_{\rm m}$.  The recent
analysis of cluster mass-to-light ratio and the mass-to-light ratio of
stars indicates that the ratio of gas to stellar mass is $\sim$10:1 in
massive clusters. There is an apparent decrease in gas mass fraction
and increase in stellar mass fraction at lower mass scales, but the
very flat surface brightness of the X-ray emission makes extension of
this result to large scale lengths uncertain. The normalization of the
scaling of mass with temperature, derived from measurements of density
and temperature profiles and assuming hydrostatic equilibrium, is
lower than predicted from simulations that do not include gas cooling
or heating and has a slightly steeper slope.  Detailed {\it Chandra}
and {\it XMM-Newton} imaging spectroscopy of several clusters show
that the form of the potential is consistent with the parameterization
of Navarro, Frenk, \& White (1997) over a factor of 100 in length
scale and that there is no evidence for a dark matter core. {\it
Chandra} X-ray images have revealed rather complex internal structures
in the central regions of some clusters, which are probably due to the
effects of mergers; however, their nature is still not completely
clear.

There are now more than 100 clusters with well-determined Fe
abundance, several with accurate values at redshifts $z \approx 0.8$,
with little or no evidence for evolution in the Fe abundance with
redshift. There is real variance in the Fe abundance from cluster to
cluster, with a trend for clusters with higher gas densities to have
higher Fe abundances. The Si, S, and Ni abundances do not follow
patterns consistent with simple sums of standard Type Ia and Type II
supernova, indicating that the origin of the elements in clusters is
different from that in the Milky Way. The Si/Fe abundance rises with
cluster mass, but the S/Fe ratio does not. The high Ni/Fe ratio
indicates the importance of Type Ia supernovae. {\it XMM-Newton}
grating spectra of the central regions of clusters have derived
precise O, Ne, Mg, and Fe abundances.  {\it XMM-Newton} CCD data are
allowing O abundances to be measured for a large number of clusters.
\end{abstract}

\section{Introduction}

Clusters of galaxies are the largest and most massive collapsed
objects in the Universe, and as such they are sensitive probes of the
history of structure formation. While first discovered in the optical
band in the 1930's (for a review, see Bahcall 1977a), in same ways the
name is a misnomer since most of the baryons and metals are in the hot
X-ray emitting intracluster medium and thus they are basically ``X-ray
objects.''  Studies of their evolution can place strong constraints on
all theories of large-scale structure and determine precise values for
many of the cosmological parameters. As opposed to galaxies, clusters
probably retain all the enriched material created in them and being
essentially closed boxes they provide an unbiased record of
nucleosynthesis in the Universe. Thus, measurement of the elemental
abundances and their evolution provide fundamental data for the origin
of the elements. The distribution of the elements in clusters reveals
how the metals were removed from stellar systems into the
intergalactic medium (IGM). Clusters should be fair samples of the
Universe, and studies of their mass and their baryon fraction reveal
the gross properties of the Universe as a whole. Since most of the
baryons are in the gaseous phase and clusters are dark matter
dominated, the detailed physics of cooling and star formation are much
less important than in galaxies. This makes clusters much more
amenable to detailed simulations than galaxies or other systems in
which star formation has been an overriding process. Detailed
measurements of their density and temperature profiles allow an
accurate determination of the dark matter profile and total mass.
While gravity is clearly dominant in massive systems, much of the
entropy of the gas in low-mass systems maybe produced by
nongravitational processes.
%XX shocks makes no sense

Clusters are luminous, extended X-ray sources and are visible out to
high redshifts with present-day technology. The virial temperature of
most groups and clusters corresponds to $kT \approx (2-100) \times
10^6$ K (velocity dispersions of 180--1200 km s$^{-1}$), and while
lower mass systems certainly exist, we usually call them
galaxies. Most of the baryons in groups and clusters of galaxies lie
in the hot X-ray emitting gas, which is in virial equilibrium with the
dark matter potential well [the ratio of gas to stellar mass is
$\sim$(2--10):1; Ettori \& Fabian 1999]. This gas is enriched in heavy
elements (Mushotzky et al. 1978) and is thus the reservoir of stellar
evolution in these systems. The presence of heavy elements is revealed
by line emission from H and He-like transitions, as well as L-shell
transitions of the abundant elements. Most clusters and groups are too
hot to have significant line emission from C or N, but all abundant
elements with $Z > 8$ (O) have strong lines from H and He-like states
in the X-ray band, and their abundances can be well determined.

Clusters of galaxies were discovered as X-ray sources in the late
1960's (see Mushotzky 2002 for a historical review), and large samples
were first obtained with the {\it Uhuru} satellite in the early 1970's
(Jones \& Forman 1978). Large samples of X-ray spectra and images were
first obtained in the late 1970's with the {\it HEAO} satellites (see
Forman \& Jones 1982 for an early review). The early 1990's brought
large samples of high-quality images with the {\it ROSAT} satellite
and good quality spectra with {\it ASCA} and {\it Beppo-SAX}.  In the
last three years there has been an enormous increase in the
capabilities of X-ray instrumentation due to the launch and operation
of {\it Chandra} and {\it XMM-Newton}. Both {\it Chandra} and {\it
XMM-Newton} can find and identify clusters out to $z > 1.2$, and their
morphologies can be clearly discerned to $z > 0.8$ (Fig. 1.1).  The
cluster temperatures can be measured to $z \approx 1.2$, and {\it
XMM-Newton} can determine their overall chemical abundances to $z
\approx 1$ with sufficiently long exposures (very recently the
temperature and abundance of a cluster at $z$ = 1.15 was measured
accurately in a 1~Ms {\it XMM-Newton} exposure; Hasinger et al. 2004).
Temperature and abundance profiles to $z \approx 0.8$ can be well
measured and large samples of X-ray selected clusters can be
derived. {\it Chandra} can observe correlated radio/X-ray structure
out to $z > 0.1$ and has discovered internal structure in
clusters. The {\it XMM-Newton} grating spectra can determine accurate
abundances for the central regions of clusters, in a model independent
fashion, for O, Ne, Mg, Fe, and Si.

\begin{figure*}[t]
\centering
\includegraphics[width=1.00\columnwidth,angle=0,clip]{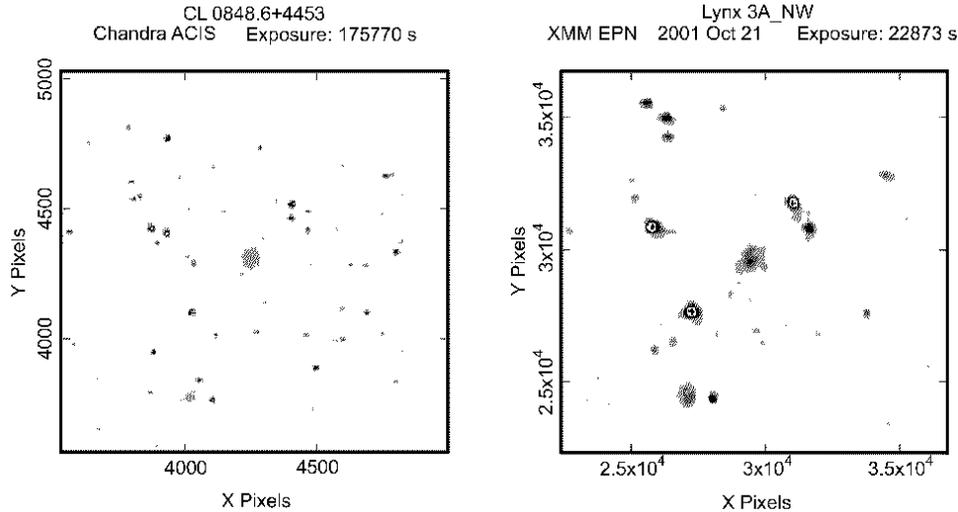}
\vskip 0pt \caption{
{\it Chandra} (left panel) and {\it XMM-Newton} (right panel) images
of the Lynx region (Stanford et al. 2001), which contains three
high-redshift clusters. The {\it Chandra} image has $\sim$170 ks and
the {\it XMM-Newton} image $\sim$20 ks exposure.}
\end{figure*}

It is virtually impossible to give a balanced review of the present
observational state of X-ray cluster research, with more than 100
papers published each year. I will not say much about those issues for
which we have had detailed talks at this meeting: cooling flows,
high-redshift clusters and evolution, X-ray data and the
Sunyaev-Zel'dovich effect, radio source interaction, X-ray selected
active galaxies in clusters, X-ray emission from groups, and detailed
comparison of masses derived from lensing and X-ray
observations. Other areas, such as the presence of nonthermal emission
and the existence of very soft components, were not discussed. Even
limiting the talk this much results in an abundance of
material. However, for the purposes of continuity, I have included
some material that overlaps with the reviews on chemical abundance
given by Renzini (2004) and on groups by Mulchaey (2004). This review
does not consider work published after February 2003.

\section{Temperature Structure of Clusters}

As discussed in detail by Evrard (2004), we now have a detailed
understanding of the formation of the dark matter structure for
clusters of galaxies. If gravity has completely controlled the
formation of structure, one predicts that the gas should be in
hydrostatic equilibrium with the vast majority of the pressure being
due to gas pressure.  If this is true, its density and temperature
structure provide a detailed measurement of the dark matter
distribution in the cluster. Recent theoretical work has also taken
into account other process such as cooling and turbulence, which can
be important.  The fundamental form of the Navarro, Frenk, \& White
(1997; hereafter NFW) dark matter potential and the assumption that
the fraction of the total mass that is in gas is constant with radius
result in a prediction, both from analytic (Komatsu \& Seljak 2001)
and numerical modeling (Loken et al.  2002), that the cluster gas
should have a declining temperature profile at a sufficiently large
distance from the center (in units of $R/R_{\rm virial}$).  The size
of the temperature drop in the outer regions is predicted to be
roughly a factor of 2 by $R/R_{\rm virial} \approx 0.5$, which is
consistent with the {\it ASCA} results of Markevitch (1998). However,
there is considerable controversy about the analysis and
interpretation of temperature profiles before {\it XMM-Newton} and
{\it Chandra}. Results from both {\it ASCA} (Kikiuchi et al. 1999;
White \& Buote 2000) and {\it Beppo-SAX} (Irwin \& Bregman 2000;
De~Grandi \& Molendi 2002), indicate either isothermal gas or a
temperature gradient in the outer regions of some ``cooling flow''
clusters. {\it XMM-Newton} is perfect for resolving this controversy,
having a much better point spread function than {\it ASCA} and much
more collecting area than {\it Beppo-SAX} and {\it Chandra}, and
having a larger field of view than {\it Chandra}. However, there is a
selection effect due to the smaller {\it XMM-Newton} field of view
than {\it ASCA}, and in order to go out to the virial radius in one
pointing one must observe clusters at $z> 0.1$.

There are several published temperature profiles from {\it XMM-Newton}
(Tamura et al. 2001; Majerowicz, Neumann, \& Reiprich 2002; Pratt \&
Arnaud 2002) and I have analyzed several other moderate redshift
clusters and others were presented at this conference (Jones et
al. 2004).  With the exception of one object (A1101S; Kaastra et
al. 2001) all the published {\it XMM-Newton} profiles are consistent
with isothermal profiles out to $R/R_{\rm virial} \approx 0.5$
(Fig. 1.2), which is in strong disagreement with the numerical and
analytic modeling. This sample of $\sim$12 objects is highly biased to
smooth, centrally condensed clusters (with the exception of Coma,
which has been known to be isothermal from the early work of Hughes et
al. 1988). The data for A2163 are consistent with a temperature drop
at even larger radii (Pratt, Arnaud, \& Aghanim 2002), but the
relatively high {\it XMM-Newton} background makes the results somewhat
uncertain. The origin of the difference between some of the {\it
Beppo-SAX}, {\it ASCA}, and {\it XMM-Newton} results is not clear. It
is possible that there is a difference between the low-$z$ systems
studied by {\it Beppo-SAX} and {\it ASCA} and the higher-redshift
systems studied by {\it XMM-Newton} and/or a selection effect in the
objects so far analyzed with {\it XMM-Newton}. While the {\it Chandra}
data do not go out to very large length scales (Allen, Schmidt, \&
Fabian 2002), analysis of 2 $z \approx 0.7$ clusters with {\it
Chandra} (Ettori \& Lombardi 2003) also show isothermal profiles.

 We must now take seriously the disagreement between theory and
observation in the temperature profiles in comparing cluster
properties with simulations. Another serious issue is the inability of
theoretical models to match the observed temperature drops in the
centers of the ``cooling flow'' clusters. The question is then, what
is the origin of the discrepancy? Several possibilities are that the
form of the theoretical potential is incorrect, that the gas
distribution is not calculated correctly, or that physics other than
gravity needs to be included.

%XX poor quality
\begin{figure*}[t]
\centering
\includegraphics[width=0.80\columnwidth,angle=0,clip]{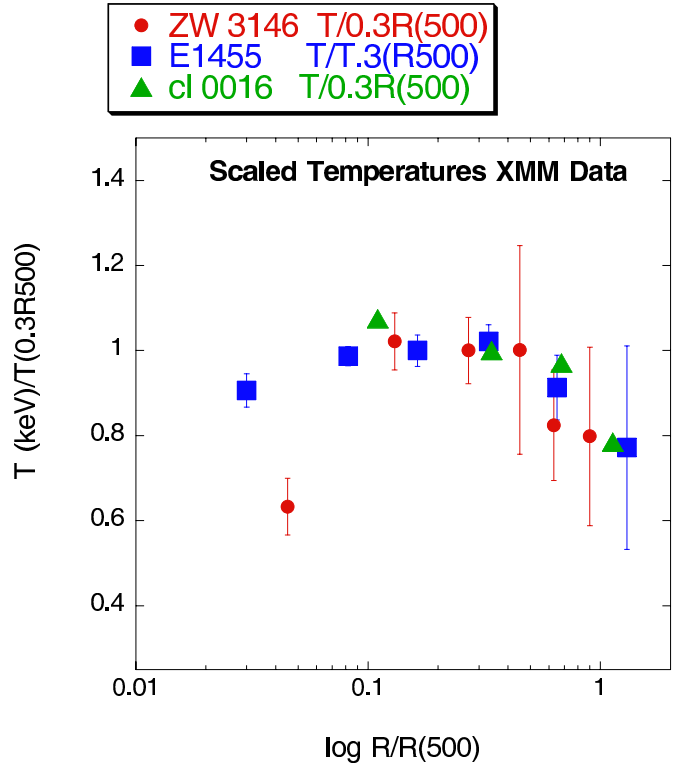}
\vskip 0pt \caption{
Normalized temperature profiles of three moderate redshift clusters
derived from {\it XMM-Newton} EPIC observations. The ratio of the
temperature in a radial bin vs. the radius in units of $R_{500}$ is
plotted.}
\end{figure*}

%\clearpage

\noindent
As shown below (\S 1.8) the form of the potential from X-ray imaging
spectroscopy agrees quite well with the NFW potential, which is
consistent with the analytic work. {\it ROSAT} and {\it XMM-Newton}
analysis of X-ray surface brightness distributions (\S 1.5) shows that
the $\beta$ model is a good description of the X-ray surface
brightness at large radii. This leaves us with the possibility that
additional physics is needed. Recent analysis of {\it Chandra} data
(cf. Markevitch et al. 2003) strongly constrains the effects of
conduction, which will tend to make isothermal spectra, while the
inclusion of cooling and heating in the theoretical models (Loken et
al. 2002) does not seem to affect the temperature profile
significantly. Thus, the origin of this severe discrepancy is not
currently known.

%XX labels, what are other three lines?
\begin{figure*}[t]
%\centering
\hspace*{-1.5cm}
\includegraphics[width=0.85\columnwidth,angle=90,clip]{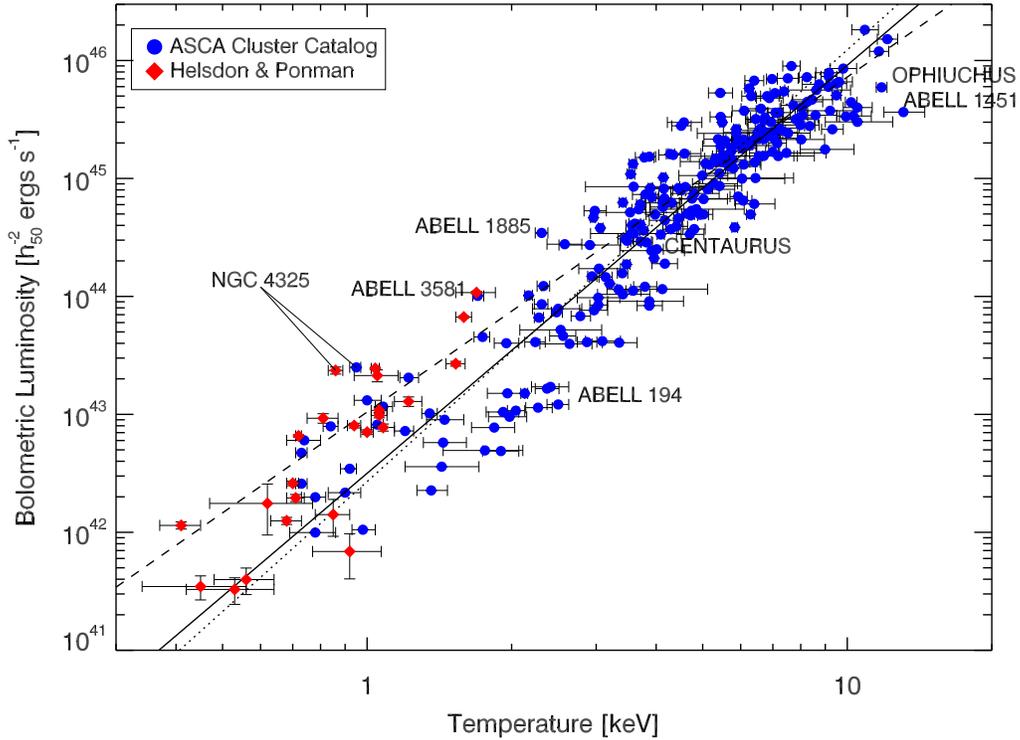}
\vskip 0pt \caption{
X-ray luminosity vs. X-ray temperature, derived from {\it ASCA} observations
of $\sim$270 clusters (Horner 2001). The best-fit $L_{\rm X}\propto T^3.4$
for the overall sample (solid line), while the best fit for clusters
of luminosity less than $2/times 10^{44}$ ergs s$^{-1}$ 
is drawn as a dotted line}
\end{figure*}

\section{Luminosity-Temperature Relation for Clusters}

As pointed out by Kaiser (1986), simple scaling relations predict that
the cluster luminosity should scale as the temperature squared
($T^2$).  To see this, note that the X-ray luminosity should scale as
the density squared times the volume times the gas emissivity, $L_{\rm
X} \propto \rho^2 V \Lambda$.  The mass of gas scales like $\sim \rho
V$, and it is assumed that the total mass $M_{\rm T}$ scales as
$M_{\rm gas}$.  Since the emissivity for bremsstrahlung, the prime
cooling mechanism in gas hotter than 2 keV, scales as $T^{0.5}$
(Sutherland \& Dopita 1993), one has $L_{\rm X} \propto M \rho
T^{0.5}$.  Finally, since, theoretically, the total mass scales as
$T^{1.5}$, one has $L_{\rm X} \propto \rho T^2$.  The other free
parameter, the average density, is related to the mass and collapse
epoch of the cluster.

It has been known for 20 years (Mushotzky 1984) that the actual
relationship between temperature and luminosity is steeper than the
simplest theoretical prediction. Recently, Horner et al. (2004) have
examined the $L_{\rm X}-T$ relation using the largest sample of
clusters to date (270 clusters taken from the {\it ASCA} database).
In this sample one finds that, over a factor of $10^4$ in luminosity,
the luminosity scales as $T^3$. As one goes to lower luminosities
there is a wider range of luminosity at a fixed temperature
(Fig. 1.3), but there is no need to change the scaling law. This
increase in variance probably explains the steeper fit at low
luminosity found by Helsdon \& Ponman (2000). This continuity is
rather strange, since at $T<2$ keV the cooling function changes sign
and scales more like $T^{-1}$, and thus the theoretical relation
between $L_{\rm X}$ and $T$ changes slope.

There have been many papers written about the origin of the
discrepancy, but the main conclusion is that it is due to the breaking
of scaling laws via the inclusion of physics other than gravity. The
same physics that helps to explain the deviation of entropy in groups,
such as heating and cooling, can also explain the slope and
normalization of the $L_{\rm X}-T$ relation (see Mulchaey 2004
and Borgani et al. 2002).  Another indication of this
scale breaking is the relative low level of evolution in the $L_{\rm
X}-T$ relation out to $z\approx 1$ (Borgani et al. 2002) which is not
what is predicted in simple theories of cluster evolution, since
objects at $z\approx 1$ are predicted to be denser and have a higher
temperature for a fixed mass.  Simple scaling predicts that $T \propto
M^{1.5} (1+z)$, and thus one predicts $L \propto T^2 (1+z)^{0.5}$ at a
fixed mass, which is not seen (but see Vikhlinin et al. 2002 for a
different opinion).

It was pointed out by Fabian et al. (1994) that high central density,
short cooling time clusters (alias ``cooling flow'' clusters) have a
considerably higher luminosity for their temperature than non-cooling
flow systems. This result is confirmed in the larger Horner et
al. (2004) sample. Markevitch (1998) removed the high-central surface
brightness central regions from these clusters and found that the
scatter in the $L_{\rm X}-T$ relationship was much reduced and the fit
was flatter than $T^3$. If the scatter in the $L_{\rm X}-T$
relationship was due to cool gas in the center of the cooling flow
clusters, one should expect that the {\it ROSAT} luminosities, which
are very sensitive to low-temperature gas, would be systematically
larger than the luminosities calculated from isothermal fits to the
{\it ASCA} data. However, Horner et al. (2004) find that the
bolometric luminosities obtained by {\it ASCA} are in very good
agreement with the {\it ROSAT} results. This indicates that the
central luminosity ``excess'' is not due to cool gas, as was
originally shown in the {\it ASCA} data for the Centaurus cluster
(Ikebe et al. 1999) and recently shown in detail by {\it XMM-Newton}
spectroscopy of many cooling flow clusters (Peterson et al.
2003). Horner et al. (2004) find that the most reasonable explanation
for the higher luminosity of the cooling flow clusters is due to their
higher central density in the core. This result is consistent with the
detailed analysis of cluster surface brightness profiles by Neumann
\& Arnaud (2001) (see \S 1.5). 
It thus seems that the scatter in the $L_{\rm X}-T$
relation at high temperatures is due to differing cluster central gas
densities, while the scatter at low temperatures is due to different
``amounts'' of additional, nongravitational physics.

%XX poor quality
\begin{figure*}[t]
\centering
\includegraphics[width=1.00\columnwidth,angle=0,clip]{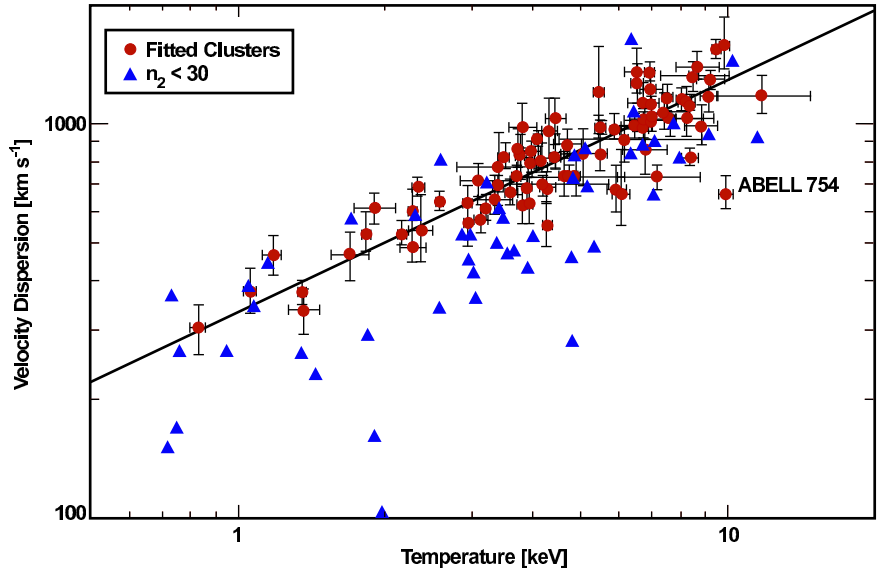}
\vskip 0pt \caption{
X-ray temperature vs. velocity dispersion, taken from Horner (2001).  The
triangles represent points with fewer than 30 galaxies per cluster. Note
that these points contribute much of the variance in the fit.}
\end{figure*}

\section{Optical Light, Velocity Dispersion, and X-ray Properties}

%XX wide range of what?  between makes no sense
It has been known since the early {\it Uhuru} results (Jones \& Forman
1978) that there is a great degree of scatter in the correlation
between cataloged optical properties, such as Abell richness, and
X-ray properties, such as luminosity and temperature. The best
correlations between optical and x-ray properties seen in the early
data were between central galaxy density and X-ray luminosity (Bahcall
1977b), and between X-ray temperature and optical velocity dispersion
(Edge \& Stewart 1991). The wide scatter is nicely illustrated in
Figure 5 of Borgani \& Guzzo (2001)which shows that the Abell counts
are only weakly related to total mass, while the x-ray luminosity is
strong correlated.

Bird, Mushotzky, \& Metzler (1995) showed that much of the scatter in
the temperature - velocity dispersion correlation was due to
undersampled optical data and velocity substructure in the
clusters. More recent optical and X-ray work (Girardi et al. 1998;
Horner et al. 2004) shows that when the velocities of a sufficient
number of galaxies in a cluster are measured (one needs more than 30
galaxies) (Fig. 1.4) there is a tight relation between velocity
dispersion and temperature of the form $\sigma
\propto T^{0.59\pm0.03}$, consistent with the work of Bird et
al. (1995) and close to the theoretical slope of 0.5. This has been
confirmed in an infrared-selected sample by Kochanek et
al. (2003). The normalization of this relation at high temperatures
agrees with theoretical work (Evrard 2004), and thus one has to
conclude that low-velocity dispersion clusters are too hot for their
dispersion, or that low-temperature clusters have too low a dispersion
for their temperature.  The fact that clusters have very small radial
velocity dispersion gradients (Biviano \& Girardi 2003) or temperature
gradients (\S 1.2) makes comparison of the average temperature and
velocity dispersion meaningful. This variation with temperature of the
velocity dispersion to temperature ratio will also change the
effective X-ray vs. optically determined mass by a factor of 50\% over
the full mass range of clusters.

Recent 2MASS work by Kochanek et al. (2003) shows that, if the
``optical'' data are handled carefully (e.g., accurate photometry,
well-defined selection criteria, observing in a red passband, etc.),
there is a strong relation between the total light in a cluster and
the X-ray temperature and luminosity (also see Yee \& Ellingson
2003). However, while the correlations are much better than in
previous work, the scatter in the relation is large, almost a factor
of 10 in light at a fixed X-ray temperature or luminosity, or,
alternatively, a factor of $\sim$2 in temperature at a fixed optical
luminosity. Thus, one expects that optical and X-ray catalogs of
clusters might differ considerably depending on where the cuts are
made. There is no evidence for either optically or X-ray quiet
clusters, but there is evidence for relatively optically or X-ray
bright objects. The nature and origin of this variance is not
understood at present, but, given the quality of modern data, this
variance seems to be real, rather than due to measurement
uncertainties. Assuming that the X-ray properties accurately trace
mass, the $K$-band light is a mass indicator accurate to 50\% (Lin,
Mohr, \& Stanford 2003). The converse test, estimating the mass from
the optical data and comparing it to the X-ray data, shows large
scatter (Yee \& Ellingson 2003), where the temperature data are taken
from the literature. If it is indeed the case that there is a large
variation in the ratio of optical light to X-ray temperature, this
indicates that there is a considerable variance in cluster
mass-to-light ratio at a fixed mass. This would be a major challenge
to structure formation theories.

\section{Surface Brightness Profiles}
It has been known since the pioneering work of Jones \& Forman (1984)
that the surface brightness profiles of most clusters can be well fit,
at large radii, by the ``isothermal'' $\beta$ model,
$S(r)=S_0(1+(r/a)^2)^{(-3\beta + 0.5)}$, with a central excess above
the $\beta$ model in cooling flow clusters. As seen in {\it ROSAT}
data for high-redshift systems (Vikhlinin, Forman, \& Jones 1999), the
$\beta$ model fits amazingly well out to the largest radii measurable
for massive clusters. The fitted values of $\beta$ are smaller for
low-mass systems (Helsdon \& Ponman 2000; Mulchaey et al. 2003), but
there are two selection effects that make the interpretation of this
result difficult. First of all, because of their low surface
brightness, the group profile hits the background at relatively small
distances from the center, and thus one does not detect low-mass
systems out to large fractions of the virial radius. This can
introduce a bias to the fitted values of $\beta$. Secondly, the
effects of the central galaxy on the surface brightness is often not
well determined from {\it ROSAT} PSPC data (Helsdon \& Ponman 2003)and
thus, frequently, the structural parameters are not well constrained.
This latter effect is not present in {\it XMM} or {\it Chandra} data.

Neumann \& Arnaud (2001) have pointed out that the surface brightness
profiles of high-temperature clusters remain self-similar as a
function of mass and redshift, as expected from cold dark matter
models (see also Vikhlinin et al. 1999). Since the conversion from
angle to distance depends on the cosmology, they have been able to
show that the change of profile with redshift is most consistent with
a $\Lambda$ dominated cosmology. The homology of the profiles is only
applicable outside of the central 100 kpc, as inside this radius there
are often large deviations from the scaling laws. However, in order to
achieve the scaling they require that the relationship of gas mass to
temperature be $M_{\rm gas} \propto T^2$, steeper than the theoretical
scaling between total mass and temperature (i.e., $M_{\rm
total}\propto T^{1.5}$).  Since the surface brightness profiles scale
according to the predicted evolution from the cold dark matter models,
the lack of evolution in the luminosity-temperature law must be a
cosmic conspiracy between the cosmological model and the change of
density with redshift. The prediction is that the emission measure of
the gas scales as $EM \propto \beta f^2_{\rm
gas}\Delta^{1.5}(1+z)^{9/2}(kT)^{0.5} h^3$, where $\Delta$ is the
overdensity of the cluster and $f_{\rm gas}$ is the fraction of mass
that is in gas (Arnaud, Aghanim, \& Neumann 2002).

There are ``single'' clusters that are not well fit by the $\beta$
model.  The most obvious example is MS 1054$-$0321 at $z$ = 0.82
(Jeltema et al.  2001), which is much more concentrated than a $\beta$
model.  This is not a function of redshift, since many clusters at
$z>0.6$ are well fit by the $\beta$ model.

\section{Mass of  Baryons and Metals and How They Are
Partitioned}

%XX shorten section title to fit

The two main baryonic components of clusters are the X-ray emitting
gas and the stars, since the total contribution from cold gas and dust
is very small. The major uncertainty in the relative baryonic
contribution is due to the uncertainty in the transformation from
light to mass for the stars.  Recent work from large optical surveys
(Bell et al. 2003) shows that the mass-to-light ratio of stars changes
as a function of galaxy but is $\sim$3.5 in the Sloan $g$ band for a
bulge-dominated population. Using this value and the mean
mass-to-light ratio of clusters $\sim$240 (Girardi et al. 2002), the
stars have $\sim$0.015 of the total mass. The gas masses have been
well determined from {\it ROSAT} data (Ettori \& Fabian 1999; Allen et
al. 2002) and scatter around $f_{\rm gas} \approx 0.16 h_{70}^{-0.5}$.
Thus, the gas-to-stellar mass ratio is $\sim$10:1, and the total
baryon fraction is almost exactly consistent with the recent {\it
WMAP} results for the Universe as a whole.  Since it is thought that
clusters are representative of the Universe as a whole, this suggests
that the vast majority of baryons in the Universe do not lie in
stars. Turning this around, one can use the baryonic fraction in
clusters as a bound on $\Omega_{\rm m}$ (White et al. 1993). The most
recent analysis using this technique finds $\Omega_{\rm m} < 0.38
h_{70}^{-0.5}$ (Allen et al. 2002), in excellent agreement with the
{\it WMAP} data. It is interesting to note that the high baryonic
fraction in clusters has been known for over 10 years and was one of
the first strong indications of a low $\Omega_{\rm m}$ Universe.
Since it is thought that the baryonic fraction in clusters should not
evolve with redshift, derivation of the baryonic abundance in high-$z$
clusters, which depends on the luminosity distance, provides a strong
constraint on cosmological parameters (Ettori \& Tozzi and Rosati
2003).

The mean metallicity of the gas in clusters is $\sim$1/3 solar (see
\S 1.10), while that of the stars may be somewhat larger. If we assume
1/2 solar abundance for the stars, than $\sim$85\% of the metals are
in the gas phase. Since all the metals are made in stars, which lie
primarily in galaxies, this implies that most of metals have either
been ejected or removed from the galaxies.  Since the stellar mass is
dominated by galaxies near $L^{\star}$, which have a mean escape
velocity, today, of $>$300 km s$^{-1}$, this implies very strong
galactic winds at high redshift. This scenario is consistent with the
results of Adelberger et al. (2003) on the high-redshift, rapidly
star-forming $U$ and $B$-band drop-out galaxies, which all have
large-velocity winds.

Analysis of the gas mass fraction in groups and clusters (Sanderson et
al. 2003) indicates that the fraction apparently drops at lower masses
by a factor of 2--3, with the reduction setting in at a mass scale
corresponding to 1--3 keV at 0.3 $R_{200}$. In addition, the stellar
mass-to-light ratio decreases by 60\% over the same mass range
(Marinoni \& Hudson 2002), and thus in groups the gas-to-stellar mass
ratio is only (1--2):1 at 0.3 $R_{200}$, considerably smaller than in
clusters. However, there is a serious problem for groups in evaluating
both the gas and stellar masses at large radii (see Fig. 10 in
Mulchaey et al. 2003), and this result should be taken with some
caution. In particular, the X-ray surface brightness distribution of
groups is often very flat, and extrapolating from 0.3 $R_{200}$ to
$R_{200}$ is rather risky.  However, if these trends are real, this
would indicate that groups are truly baryon poor, that the baryons
have been pushed out of the group, or that the gas has been puffed
up. If the gas has been puffed up, this is consistent with the
somewhat high temperatures of groups compared to their optical galaxy
velocity dispersions, indicative of extra heat deposited in the gas,
which both heats it and ``puffs'' it up (see discussion in the review
by Mulchaey 2004).

\section{Mass Scaling Laws}

Detailed theoretical work has verified that clusters should satisfy the virial 
theorem, and thus their mass should scale as $M \propto T R$, with $R \propto 
T^{1/2}$, and thus $(1+z)M^{2/3} \propto T$ (e.g., Eke, Navarro, 
\& Frenk 1998), with the normalization being set by theory and the value of
the cosmological parameters (Evrard 2003). The first test of this
relation (Horner, Mushotzky, \& Scharf 1999) found a scaling that was
somewhat steeper, with $M \propto T^{1.7}$, and a normalization that
was 40\% lower than predicted. Finoguenov, Reiprich, \& B\"{o}hringer
(2001) and Reiprich \& B\"{o}hringer (2002) have confirmed these
results with more uniform samples, and higher quality, spatially
resolved spectra. Recent {\it Chandra} results (Allen, Schmidt, \&
Fabian 2001) are also consistent with the Horner et al.  (1999)
finding. {\it XMM-Newton} data for A1413 (Pratt \& Arnaud 2002) show
that the normalization scaling is not only violated by the sample, but
by individual objects.  The normalization in the Reiprich \&
B\"{o}hringer (2002) sample agrees with theoretical expectations at
the high-mass end. This indicates that lower-temperature clusters are
less massive than expected on the basis of their temperature,
consistent with the trend seen in the velocity dispersion-temperature
relation.  Recently, it has been pointed out (Shimizu et al. 2003)
that the combination of the scaling of mass by $M \propto T^{1.7}$ and
the gas mass fraction scaling as $T^{1/3}$ (a reasonable fit to the
Sanderson et al. 2003 data) can reproduce the observed $L_{\rm X}
\propto T^3$ relationship. Theoretical calculations that include the
effects of cooling (Thomas et al. 2002) seem to be consistent with the
lower normalization, but so far the slope difference has not been
explained.

\section{Form of the Potential}

As discussed extensively in this conference, the form of the potential
in clusters should be determined by the distribution of dark matter.
Recent numerical work seems to validate the NFW potential, and much
has been made of the fact that low-mass and low-surface brightness
galaxies do not seem to follow this form in their central
regions. Recent {\it Chandra} and {\it XMM-Newton} observations (Allen
et al. 2002; Arabadjis, Bautz, \& Garmire 2002; Pratt \& Arnaud 2002)
have been able to determine extremely accurate mass profiles via
spatially resolved X-ray spectroscopy and the assumption of
hydrostatic equilibrium. Perhaps the best documented of these examples
are the {\it Chandra} data for Abell 2029 (Lewis, Buote, \& Stocke
2003), in which the profile is determined over a factor of 100 in
length scale, from 0.001--0.1 characteristic lengths of the NFW
profile, with essentially no deviation from the NFW prediction. This
striking result is also seem in other {\it Chandra} results in the
cores of clusters. The data show that the central regions of clusters
tend to have rather steep density profiles in the innermost radii,
indicating that whatever causes the deviation of the form of the
potential in dwarf galaxies does not occur in clusters. This results
strongly constrains interacting dark matter models (Bautz \& Arabadjis
2004).  A survey of {\it Chandra} central mass profiles is made
somewhat difficult because of the possibly complex nature of the IGM
in the central regions of many clusters, and the exact slope and
normalization of the mass depends on the details of the thermal model
used. However, if the data are of sufficiently high signal-to-noise
ratio, the form of the mass profile can be determined precisely. I
anticipate quite a few exciting new results in this area; preliminary
results, presented in several conferences, indicate a predominance of
steep mass profiles with slopes close to the NFW level, but with some
scatter.

\begin{figure*}[t]
\centering
\includegraphics[width=1.00\columnwidth,angle=0,clip]{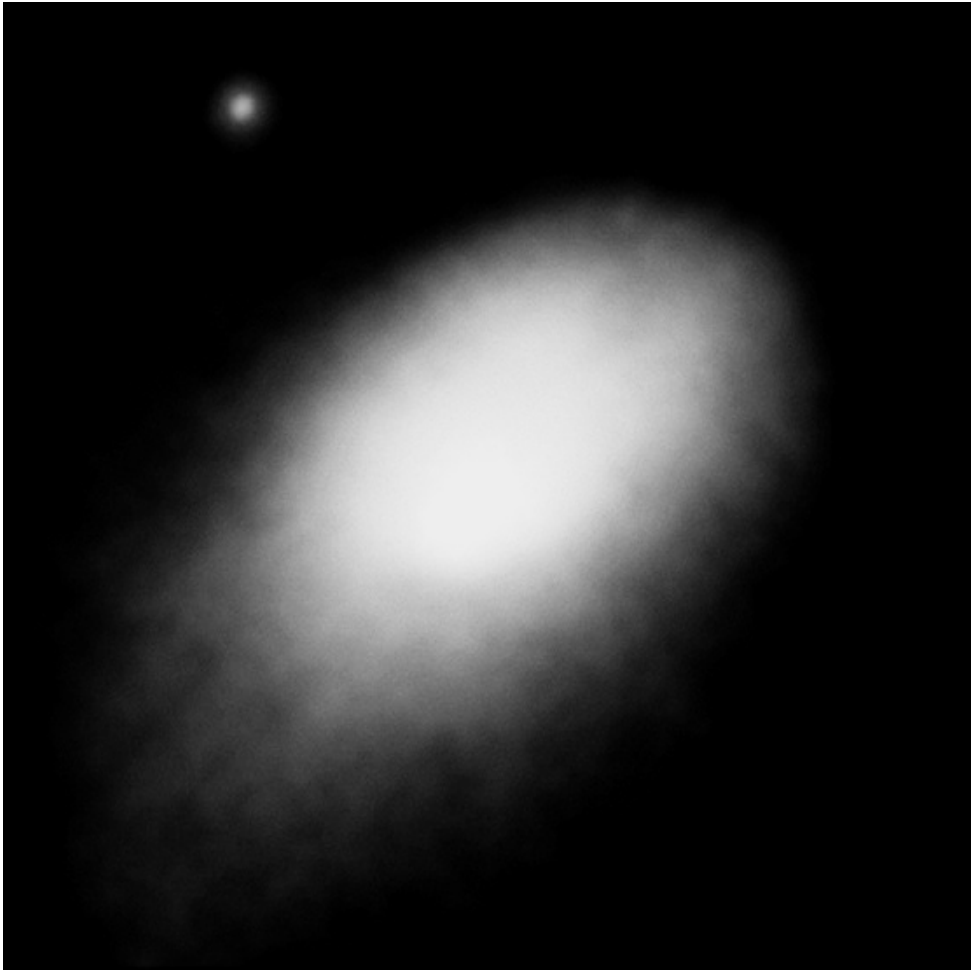}
\vskip 0pt \caption{
The {\it Chandra} image of a cold front in Abell 2142 (courtesy of the
{\it Chandra} Science Center). Note the very sharp boundary to the
north, which is not an artifact of the image processing.}
\end{figure*}
%XX scale?

\section{Merges, Structures, etc.}
The early {\it Einstein} Observatory images of clusters (Henry et
al. 1979) showed that a substantial fraction of the X-ray images were
not simple, round systems, but often complex in form and sometimes
even double. This observation is consistent with the idea that
clusters form in a hierarchical fashion via mergers, and that the
complex systems are in the process of merging. The fact that mergers
are actually occurring, rather than the complex structures in the
images being simply projection effects, was indicated by complexity in
the temperature structure of many of these systems shown by the {\it
ROSAT} (Briel \& Henry 1994) and {\it ASCA} (Markevitch 1996)
data. The details of the nature of this process have had to wait until
the precise {\it Chandra} spectral images showed the full range of
complexity. While ``textbook'' examples of merger shocks have been
seen (e.g., 1E~0657$-$56; Markevitch et al. 2002), many of the objects
show only subtle temperature variations (e.g., Sun et al. 2002). These
variations have only shown up in the most recent, very high-resolution
numerical simulations, indicating the non-intuitive nature of these
data.

The recent spectacular {\it XMM-Newton} temperature image of the
Perseus cluster (Churazov et al. 2003) illustrates the wealth of
detail that is now possible to obtain. It is interesting that these
spectral images do not show the numerous ``cold spots'' that are
predicted in cluster simulations that include cooling (Motl et
al. 2004). The ability to obtain spectral images has also revealed
``hidden mergers.'' Both the Coma and Ophiuchus clusters, the hottest
nearby systems, show smooth, regular X-ray images; however, X-ray
temperature maps show strong spatial variations (Arnaud et al.  2001;
Watanabe et al.  2001). So far the data on abundance variations in the
mergers is sparse, but the abundances seem uniform, within errors, in
Coma and may vary by less than a factor of $\sim$2 in Ophiuchus. It
seems as if many of the large-scale length, non-cooling flow clusters
are recent mergers.

One of the surprises of the {\it Chandra} data was the discovery of
surface brightness discontinuities in the surface brightness --- the
so-called cold fronts (Fig. 1.5; Vikhlinin \& Markevitch 2002). These
cold fronts are apparently contact discontinuities, across which the
pressure is smoothly varying but the density and temperature change
discontinuously by factors of $\sim$2. They can occur in ``pure
hydro'' numerical simulations (Bialek, Evrard, \& Mohr 2002). Their
relative frequency is a indication of the merger rate. However, the
details (e.g., temperature drop, size of region, etc.) and their
relation to merger dynamics are not certain (Fujita et al. 2002). The
stability of the cold fronts, their sizes, and shapes are indications
of the strength of the magnetic field, velocity vector of the merger,
and the amount of turbulence (Mazzota, Fusco-Femiano, \& Vikhlinin
2002) in the cluster gas. It is clear that there is much to learn from
further studies of these unexpected structures, but they already
confirm that the gas is usually not strongly shocked, nor highly
turbulent.

\section{Abundances}
As indicated above, most of the metals in the cluster lie in the hot,
X-ray emitting gas. Thus, in order to understand the formation and
evolution of the elements one must determine accurate abundances, the
abundance distribution in the gas, and its evolution with cosmic
time. Before giving the results it is important to remind the reader
that the measurement of abundances in the cluster gas via X-ray
imaging spectroscopy is a robust process. Most of the baryons and
metals are in the hot gas, and the spectral signature of the heavy
elements are relatively strong H and He-like lines (Fig. 1.6). This is
a well-understood emission mechanism, with little or no radiative
transfer difficulties. Because of the high temperature and short
spallation times, dust is destroyed rapidly and thus is not a problem.
The deep potential well captures an integrated record of all the
metals produced, and thus the derived abundances are true averages of
the metal production process. All the abundant elements from oxygen to
nickel can have their abundances determined.  Direct measurement of
the electron temperature from the form of the continuum and from
ratios of H to He-like lines ensures small systematic errors in the
abundances. The strongest lines in the spectrum of hot clusters are
due to Fe and Si, followed by O, S and Ni. The emission from Ne and Mg
is blended with Fe L-shell lines from Fe~XVII--XXIV at the resolution
of X-ray CCDs, and the lines from Ca and Ar are weak. With present-day
technology, one can measure Fe to $z\approx 1$ and Si to $z \approx
0.4$, and can thus obtain a true measure of the metal formation
mechanism and its evolution. For much of the rationale and background
for cluster abundance measurements, see Renzini's (2004) review in
this volume.

\begin{figure*}[t]
%\centering
\hspace*{-0.5cm}
\includegraphics[width=1.05\columnwidth,angle=0,clip]{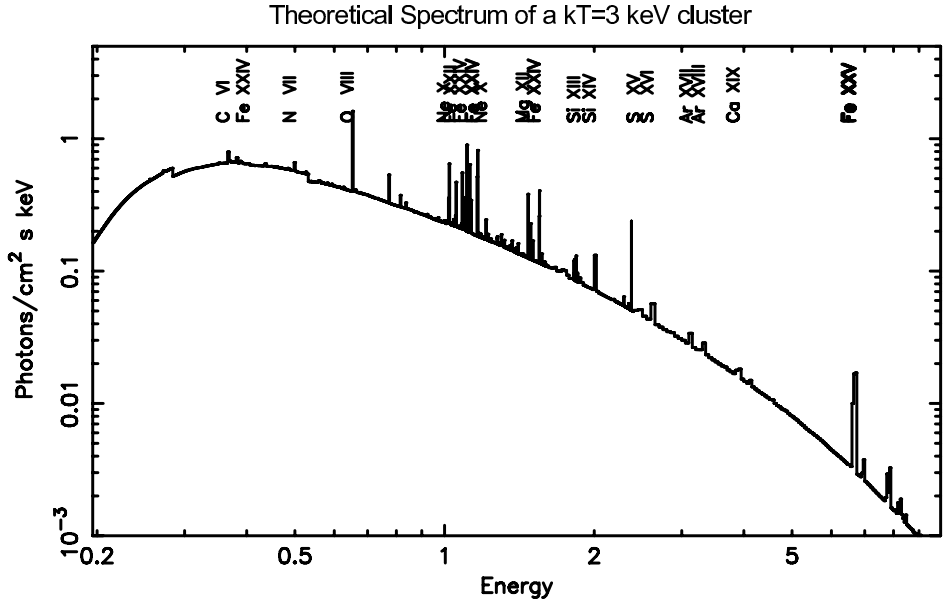}
\vskip 0pt \caption{
Theoretical spectrum of an isothermal plasma with $kT \approx 3$ keV, with
many of the strong transitions labeled; however, many of the transitions from
the same ion (e.g., multiple lines from Fe~XXIV) have been suppressed. Note
the strong lines from all the abundant elements. The data are plotted in the
usual way for X-ray astronomers, photons cm$^{-2}$ s$^{-1}$ keV$^{-1}$ vs. keV,
which emphasizes the dynamic range of X-ray spectroscopy.}
\end{figure*}
%XXX labels!!, botoom right too, upper unnecessary

\begin{figure*}[t]
%\centering
\hspace*{-1.0cm}
\includegraphics[width=0.80\columnwidth,angle=90,clip]{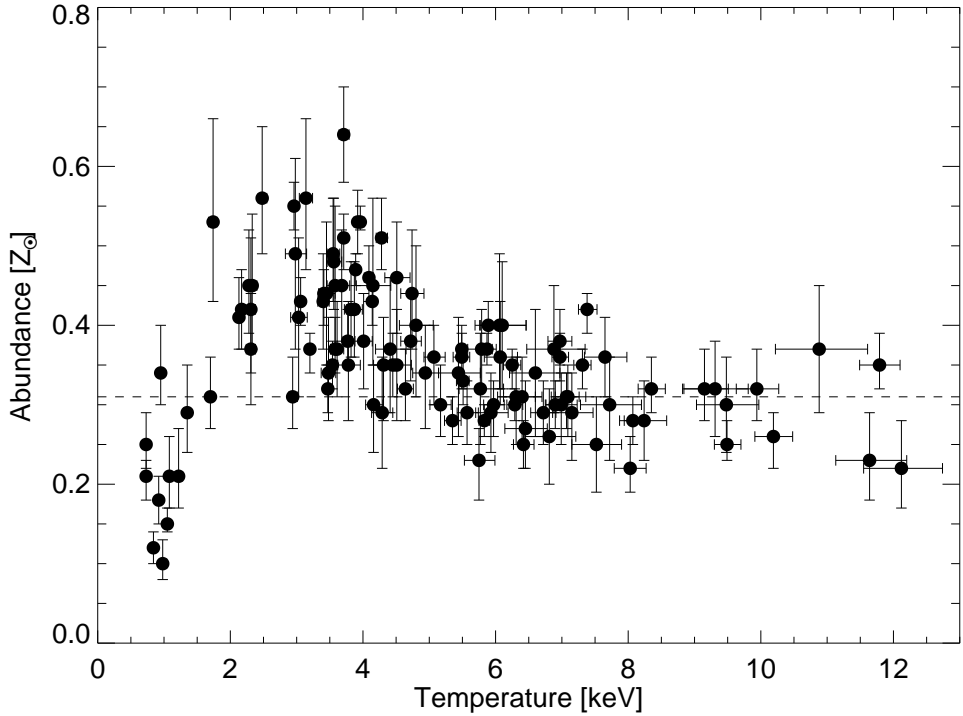}
\vskip 0pt \caption{
Fe abundance vs. temperature for the {\it ASCA} cluster database
(Horner 2001). The dashed line shows the average for clusters with
temperatures greater than 5 keV. Notice that there is little variation
at high temperatures, but there is a systematic rise in abundance and
then fall at low temperatures.}
\end{figure*}

Recently (Baumgartner et al. 2004; Horner et al. 2004), a uniform
analysis of the {\it ASCA} cluster database of 270 clusters has been
performed, which updates previous work (e.g., Mushotzky et al. 1996;
Fukazawa et al. 2000) on cluster abundances. Horner et al. (2004) and
Baumgartner et al. (2004) measure the average cluster Fe, Si, S, and
Ni abundances with no spatial information.  They find (Fig. 1.7) that
the Fe abundance is not the same for all clusters, but shows a small
spread of a factor of $\sim$2. In agreement with Fabian et al. (1994),
the cooling flow clusters show, on average, a higher Fe abundance.
There is no evidence for any evolution in the Fe abundance out to $z
\approx 0.5$ on the basis of {\it ASCA} data.  Recent {\it XMM-Newton}
and {\it Chandra} results (Jones et al.  2004; Mushotzky, private
communication) show no evolution in the Fe abundance to $z \approx
0.8$. This lack of evolution indicates that the metals are created at
$z > 1.3$ for a $\Lambda$ cosmology ( I have added in the lifetime of
the A stars that would be visible for the massive amount of star
formation necessary to produce the observed metals). Since the vast
majority of the metals are in the gas, the rate of specific star
formation (e.g., the rate per unit visible stars) would have to be
enormous to produce the elements if it were to occur at $z<2$. The Fe
abundance is weakly correlated with the temperature, reaching a
maximum at $kT \approx 2-3$ keV, but is more or less constant for
$kT>4.5$ keV clusters. Given the accuracy of recent plasma codes, the
``peak'' in abundance, which occurs at a temperature range where both
Fe L and K lines contribute to the abundance determination, is almost
certainly a real effect. The physical origin of the variance in Fe
abundance and the trends are unknown. However, since there are trends
in the apparent ratio of starlight to gas (\S 1.6), this may be the
cause. Further progress in this area requires an enhancement of the
original work of Arnaud et al. (1992), which found a correlation
between the light in elliptical galaxies and the total mass of Fe in
the cluster.

The distribution of the elements in a cluster determines the total
amount of material and gives clues as to how the material was
deposited in the IGM.  The previous generation of X-ray satellites
({\it ASCA}, {\it ROSAT}, and {\it Beppo-SAX}) derived abundance
profiles of Fe in $>$20 clusters (Finoguenov, David, \& Ponman 2000;
Irwin \& Bregman 2000; White \& Buote 2000; De~Grandi et al. 2003).
However, these results did not always agree (different analysis of the
same data and comparison of data from the same object from different
satellites produced different results). There was a tendency for
cooling flow clusters to have high central Fe abundances and larger
total abundances, suggesting a different origin of IGM enrichment in
the central regions, the effects of mixing by mergers on the Fe
abundance profile, or a physical difference in the origin of the
metals in cooling flow clusters.

{\it XMM-Newton} and {\it Chandra} data have much smaller systematic
errors and much better signal-to-noise ratios than the data from the
earlier observatories.  Early results are available for $\sim$15
systems --- most are isochemical at large radii, with several having
gradients in the central 100 kpc. The {\it Chandra} and {\it
XMM-Newton} data are well resolved and show that the abundance
gradients are quite concentrated toward the center (cf. David et
al. 2001; Tamura et al. 2001).  For a few objects the profiles reach
to near the virial radius (e.g., Zw~3146, Cl~0016 (Mushotzky
priv. comm.) , and A~1835; Majerowicz et al. 2002), two of which
(Zw~3156 and A~1835) are massive cooling flows do not show abundance
gradients outside ~100 kpc. Numerical evaluation of the observed Fe
abundance gradients (De~Grandi et al. 2003) shows that most of the
variation in the average Fe abundance between the cooling flow and
non-cooling flow clusters is not due to differences in the Fe
gradients. The ``excess'' amount of Fe in the central regions seen in
the cooling flow systems is correlated with the presence of a cD
galaxy, and the mass of "excess" Fe is roughly consistent with its
being produced in the stars in the central cD galaxy. This is rather
unexpected, since isolated elliptical galaxies have only $\sim$1/5 of
the Fe that should have been produced by the stars (Awaki et
al. 1994).

The fact that gradients do not dominate the average abundance allows a
direct interpretation of the {\it ASCA} average abundances.  The {\it
ASCA} database of $\sim$270 X-ray spectra allow determination of
average Fe, Si , S, and Ni in clusters of galaxies (Baumgartner et al.
2004). However, the signal-to-noise ratio for most of the individual
clusters is not adequate to derive robust S or Ni abundances, and
20--40 objects in each temperature bin must be added together to
derive average values and their variation with temperature. Since
cluster mass is directly related to the temperature and line strength
is also directly connected to temperature, this is the natural space
for averaging. As originally pointed out by Fukazawa et al. (2000), as
$T$ increases, Si/Fe increases. However, the new data show that S
remains roughly constant versus temperature. Baumgartner et al. (2004)
also find that the Ni/Fe ratio is approximately 3 times solar.  While
these are very surprising results, they are similar to previous
analysis of smaller {\it ASCA} and {\it XMM-Newton} data sets.  The
S/Fe, Si/Fe, and Ni/Fe ratios depend on the relative abundance of the
types of supernovae (SNe). Type Ia SNe produce mostly Fe and Ni, while
Type II SNe produce a wide range of elements but large ratios of the
$\alpha$ elements (O, Ne, and Si) to Fe. Si and S are produced via
very similar mechanisms, and at first sight it is hard to understand
how they could have different abundance patterns. In addition, in the
Milky Way, S almost always directly tracks Si. The fact that both
Si/Fe and S/Fe drop as Fe increases shows that there is indeed a
difference in the mechanisms producing the metals as a function of
mass scale. It seems rather unexpected that the ratio of Type II to
Type Ia SNe in the stars that live in cluster galaxies should change
with the mass scale of the cluster. However, the high Ni/Fe ratio
indicates that Type Ia SNe are important in the production of Fe, at
least in the central regions of clusters (Dupke \& White 2000), and
this high ratio does not allow a simple variation in SN type with
cluster mass to readily explain the abundances patterns seen in the
ASCA data.

{\it XMM-Newton} data allow the measurement of O abundances for a
reasonable sample of objects for the first time.  The best sample
published to date is based on the high-resolution RGS data (Peterson
et al. 2003). They find that the O/Fe ratio varies by a factor of
$\sim$2 from cluster to cluster, with no apparent correlation with
temperature. Analysis of {\it XMM-Newton} CCD data taken over a larger
scale (the RGS data sample only the central $1^{\prime}-2^{\prime}$ of
the cluster) confirm this variance. As noted in Gibson, Loewenstein,
\& Mushotzky (1997), the elemental abundance ratios averaged over the
cluster do not agree with any simple ratio of Type Ia to Type II
SNe. However, it is clear that over 90\% of the O, Ne, and Mg must
originate in Type II SNe.

The new {\it XMM-Newton} O abundances further strengthen this
conclusion.  However, some of the difficulties may be caused by
differential abundance gradients of different elements. There are
strong indications from {\it ASCA} data (Fukazawa et al. 2000;
Finoguenov et al. 2001) that the Fe/Si ratio rises in the cluster
centers, consistent with the cD galaxy being a source of Fe-rich
material, probably due to Type Is SNe.  However, the new {\it
XMM-Newton} data show that O does not follow this pattern. It is clear
that more work is necessary with larger samples and abundance profiles
before we can obtain a clear picture of the metal enrichment process
in clusters.

\section{Conclusion}
The progress in this field in the last 10 years has been amazing. The
X-ray properties of objects at redshift $z \approx 0.8$ are routinely
measured, and clusters are now X-ray detected at $z>1.15$. The use of
clusters for cosmology, an area covered in the volume by Freedman
(2004), is exploding. The physics of clusters and groups holds the key
to understanding the origin and evolution of structure and the origin
of the elements. It was the cluster data that first showed that most
of the baryons and metals in the Universe are in the hot phase, and
that the baryonic Universe, as seen by our eyes, is only a shadow of
the real Universe.  In the next few years we will continue to obtain
vast amounts of new data from {\it Chandra} and {\it XMM-Newton}, and
much of the present observations will be analyzed, interpreted, and
new patterns found. There are over 400 {\it Chandra} and {\it
XMM-Newton} observations of clusters and groups in the database so
far, with many more to be observed over the lifetimes of these
telescopes.  I anticipate many major new discoveries based on these
instruments.  Furthermore, the launch of {\it Astro-E2} in 2005 will
allow detailed measurements of cluster turbulence, accurate abundances
of many elements outside the cluster cores, and direct measures of the
thermodynamic properties of the gas.

 The field has benefited enormously from the synergistic interaction
of theory and observation. Most theorists and observers are now aware
of the major issues and the current observational capabilities.
Looking beyond the next few years, I anticipate that a major new X-ray
survey, perhaps 30 times better than {\it ROSAT}, will fly, producing
an extremely large and uniform cluster catalog complete out to $z
\approx 0.7$.  In the more distant future, the {\it Constellation-X}
mission will provide precision temperatures and abundances out to the
highest redshifts that clusters exist.

\vspace{0.3cm}
{\bf Acknowledgements}.  I would like to thank my long-time
collaborators and students at Goddard for their major contribution to
this work: Keith Arnaud, Wayne Baumgartner, Don Horner, Mike
Loewenstein, and John Mulchaey. I would like to thank the {\it
Chandra} and {\it XMM-Newton} projects for their major efforts in
developing, launching, and operating these amazing instruments. I
would also like to thank the {\it ASCA} team for their pioneering
efforts in the first X-ray imaging spectroscopy mission. I thank
M. Arnaud and D. Neumann for communicating results ahead of
publication. I also thank the organizers, especially John Mulchaey,
for an exciting and stimulating meeting.

\end{document}